\documentclass[12pt,a4paper]{article}

\usepackage{graphicx}
\usepackage[sf,SF]{subfigure}
\usepackage[format=plain,labelfont=bf,justification=RaggedRight,font=small,labelsep=space]{caption}
\usepackage{comment}
\usepackage{amsmath}

\usepackage{geometry}
\usepackage{palatino}           
\usepackage{pxfonts}            
\usepackage{amsmath}
\usepackage{pxfonts}

\renewenvironment{abstract}{\begin{center}\begin{minipage}[c]{0.925\textwidth}\linespread{1.5}\selectfont\begin{center}{\bf Abstract}\end{center}}
                           {\end{minipage}\vspace*{\baselineskip}\end{center}}

\selectfont

\usepackage[compact]{titlesec}
\titleformat*{\section}{\large\bf}
\titlespacing{\section}{0pt}{*0}{*0}
\titlespacing{\subsection}{0pt}{*0}{*0}
\titlespacing{\subsubsection}{0pt}{*0}{*0}

\usepackage{cite} 

\renewenvironment{thebibliography}[1]
{ \begin{oldthebibliography}{#1}
    \linespread{1.1}\selectfont
    \setlength{\labelsep}{5mm}
    \setlength{\parskip}{0\baselineskip}
    \setlength{\itemsep}{0.2\baselineskip}
}
{ \end{oldthebibliography} }

\usepackage[usenames]{color}

\usepackage{xspace}
\usepackage{ulem} 
\newcommand{\etal}{\textit{et al.}\xspace}
\newcommand{\ie}{\textit{i.e.}\xspace}
\newcommand{\eg}{\textit{e.g.}\xspace}

\begin{document}

\title{\vspace*{-\baselineskip}\linespread{1.1}\selectfont On the Propagation of Slip Fronts at Frictional Interfaces}
\author{
  {\large David S. Kammer, Vladislav A. Yastrebov, Peter Spijker,}\\[-2ex]
  {\large and Jean-Fran{\c c}ois Molinari}\\[1ex]
  {\normalsize Computational Solid Mechanics Laboratory}\\[-3ex]
  {\normalsize Ecole Polytechnique F{\'e}d{\'e}rale de Lausanne}\\[-3ex]
  {\normalsize EPFL, LSMS, IIC-ENAC, IMX-STI}\\[-3ex]
  {\normalsize Station 18, 1015, Lausanne, Switzerland}\\[1ex]
}
\date{\normalsize\today}

\maketitle

\begin{abstract}
  The dynamic initiation of sliding at planar interfaces between
  deformable and rigid solids is studied with particular focus on the
  speed of the slip front. Recent experimental results showed a close
  relation between this speed and the local ratio of shear to normal
  stress measured before slip occurs (static stress ratio). Using a
  two-dimensional finite element model, we demonstrate, however, that
  fronts propagating in different directions do not have the same
  dynamics under similar stress conditions. A lack of correlation is
  also observed between accelerating and decelerating slip
  fronts. These effects cannot be entirely associated with static
  local stresses but call for a dynamic description. Considering a
  dynamic stress ratio (measured in front of the slip tip) instead of
  a static one reduces the above-mentioned inconsistencies. However,
  the effects of the direction and acceleration are still present. To
  overcome this we propose an energetic criterion that uniquely
  associates, independently on the direction of propagation and its
  acceleration, the slip front velocity with the relative rise of the
  energy density at the slip tip.
\end{abstract}

\newpage

\section*{Introduction}

Many aspects in engineering, technology and science concerning
friction have impact on our daily lives~\cite{book:blau:2009}. As such
frictional motion has been studied for centuries, but a complete
physical understanding of friction is still lacking. For instance, the
transition from stick to slip (the onset of dynamic sliding) is not
well understood.  Nevertheless, the initiation of dynamic sliding is
an important aspect in many areas of science including fracture
mechanics~\cite{xia:2004,coker:2005} and
seismology~\cite{heaton:1990,ben-zion:2001,book:scholz:2002}.

The onset of dynamic sliding is often globally perceived as a uniform
transition from sticking to sliding. In reality, however, it is a much
more complex phenomenon. The shear stress distribution at an interface
is generally nonuniform and reaches therefore the shear strength only
at a narrow zone from which it might cause interface rupture.  The
repetition of such local slip events results in global sliding and
provides a possible explanation of stick-slip behavior that is
consistent with recent experiments, which showed that global sliding
is preceded by local slip propagating over parts of the contact
interface~\cite{rubinstein:2004,rubinstein:2007}. As shown
in~\cite{rubinstein:2007}, these repeating precursors increase
continuously their zone of propagation until a last precursor breaks
the entire interface and causes global sliding. The propagation speed
of interface ruptures was observed to range from
slow~\cite{baumberger:2002,rubinstein:2004,rubinstein:2007} to
supersonic~\cite{coker:2005}. Moreover, the front speed of a single
slip event can change along the propagation
path~\cite{rubinstein:2004}. By studying the stress field close to the
interface, Ben-David \etal~\cite{ben-david:2010a} observed
experimentally that the rupture velocity of the detachment front is
coupled to the local ratio of shear stress $\tau_{s}$ to normal stress
$\sigma_{s}$ measured before slip initiation.

Recently, numerical
investigations~\cite{braun:2009,maegawa:2010,bouchbinder:2011,tromborg:2011,amundsen:2012}
reproduced the general features of the experimental results
of~\cite{rubinstein:2007,ben-david:2010a} using simple spring-block
models. In this Letter we study numerically the initial slip event
using a finite element (FE) method (see
also~\cite{diBartolomeo:2010}), allowing us to access detailed
information on the onset of dynamical sliding and to re-examine the
hypothesis of Ben-David \etal~\cite{ben-david:2010a} on the
correlation between the slip front speed and $\tau_s / \sigma_s$. The
advantage of the FE method over the above-mentioned discrete
techniques is the ability to reproduce correctly the mechanical
behavior of continua (\eg isotropy, elasticity).

\section*{Model set-up}

\begin{figure}
\begin{center}
  \includegraphics[width=0.7\textwidth]{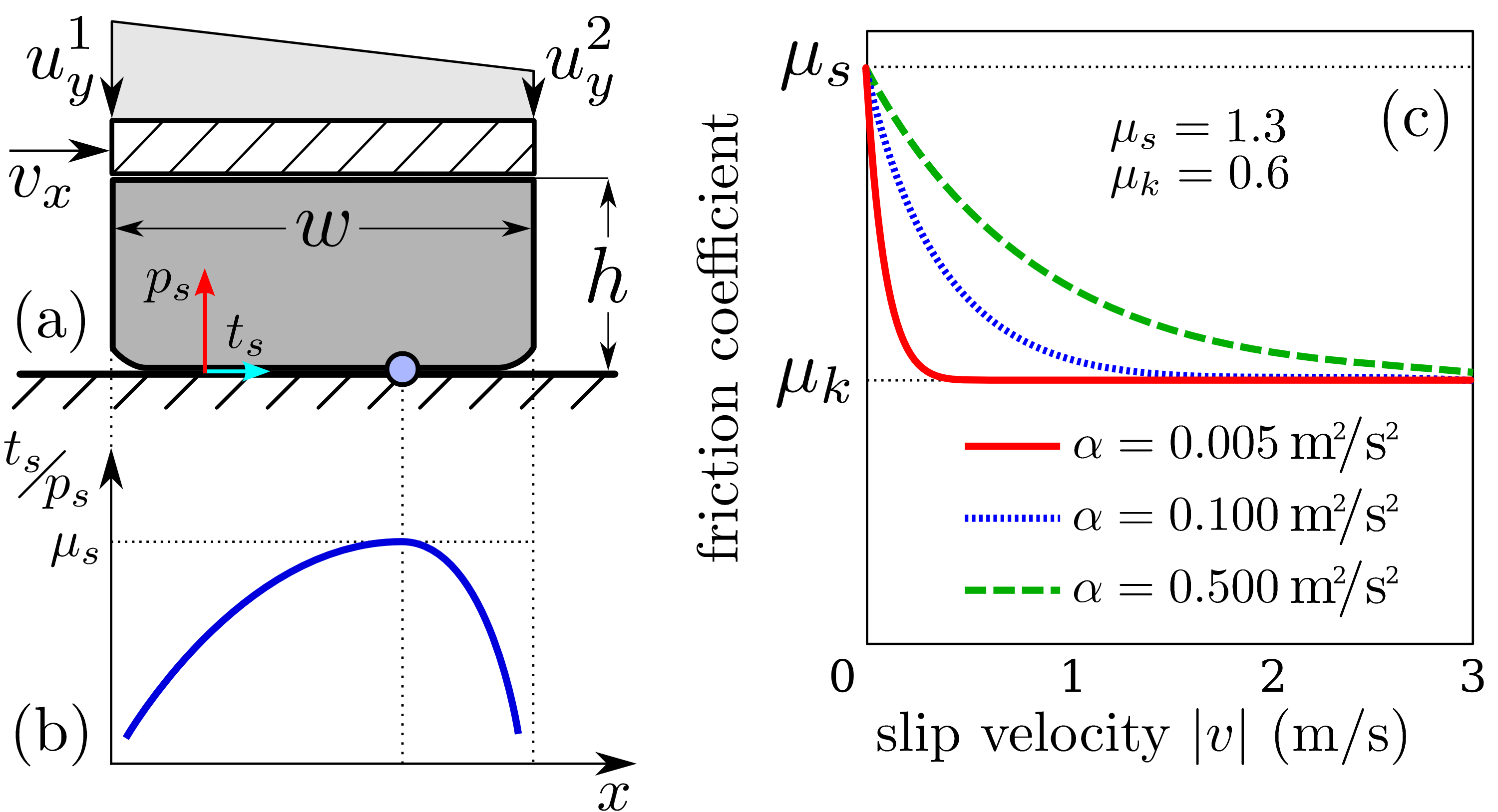}
  \caption{Two-dimensional set-up of the problem: (a) a thin
    rectangular plate in contact with a rigid plane is loaded on the
    top by a linearly distributed imposed displacement $u_y$ and a
    uniform velocity $v_x$; (b) the nonuniform distribution of shear
    to normal tractions ($t_s$ and $p_s$, respectively) at the
    interface causes a first slip nucleation far from the edges
    [$t_s/p_s>\mu_{s}$ marked by a circle in (a)]; (c) the change of the
    friction coefficient with respect to the material slip velocity
    $v$ is governed by the parameter $\alpha$ [see Eq.~\ref{eq:fric_law}]}
  \label{fig:setup}
\end{center}
\end{figure}

The two-dimensional system under consideration consists of a
rectangular isotropic elastic plate ($w=200\,$mm, $h=100\,$mm) in
contact with a rigid plane [see Fig.~\ref{fig:setup}(a)]. The corners
of the plate are rounded to avoid stress singularities at the
edges. To study this system we use a FE method with an explicit
Newmark-$\beta$ integration scheme~\cite{book:belytschko:2000} in
plane stress incorporating an energy conserving contact algorithm. The
material properties [see Table~\ref{tab:material}] correspond to
polymethylmethacrylate glass (PMMA) which was also used in the
experiments~\cite{ben-david:2010a}. We employ Rayleigh
damping~\cite{book:rayleigh:1945,caughey:1960} with mass and
stiffness proportionality coefficient of $0$ and $0.1\,\mu$s,
respectively. The deformable solid is discretized by regular
quadrilateral elements (with element side ranging for different meshes
from $0.67\,$mm to $2\,$mm) interpolating the displacement field
linearly.

A linearly distributed vertical displacement ($u_y^1=0.37\,$mm,
$u_y^2=0.037\,$mm) is imposed at the top of the plate [see
Fig.~\ref{fig:setup}(a)]. This loading is, after reaching equilibrium,
complemented by applying a uniform horizontal velocity $v_x=10^{-6}
c_{\mbox{\tiny{L}}}$, where $c_L$ is the longitudinal wave speed in
the deformable solid. The small value of the applied velocity ensures
quasi-static loading conditions, similar to the
experiments~\cite{ben-david:2010a}. The resulting stress distribution
at the interface is nonuniform. Fig.~\ref{fig:setup}(b) is a schematic
depiction of the ratio of the local tangential traction $t_s$ to the
contact pressure $p_s$. These tractions (denoted with a subscript $s$)
are measured at the moment preceding interface rupture and are
referred to hereafter as static.  The imposed loading conditions
ensure a spontaneous nucleation of the first slip event inside the
contact interface far from the edges [circle in
Fig.~\ref{fig:setup}(a)], because this is where the non-symmetric
stress distribution reaches a critical value $t_s/p_s>\mu_s$, see
Fig.~\ref{fig:setup}(b). In the stick state, the tangential resistance
of the interface is assumed to be proportional to the contact pressure
$p$ with a coefficient $\mu_s$. As for the slip state, this
coefficient of proportionality $\mu$ is determined by the velocity
($v$) weakening friction law [see Fig.~\ref{fig:setup}(c)]
\begin{equation}
\mu = \mu_k + (\mu_s-\mu_k) \exp\left(-|v|\,\sqrt{(\mu_s-\mu_k)/\alpha}\right),
\label{eq:fric_law}
\end{equation}
which ensures a smooth transition from the static $\mu_s$ to the
kinetic $\mu_k$ friction coefficient governed by the transition
parameter $\alpha$.  The parameters of the friction law are as well
presented in Table~\ref{tab:material}. The local $\mu_s$ corresponds
to experimental results as reported in Fig. 4(a-b)
in~\cite{ben-david:2011} and is considerably higher than the global
static coefficient of friction. An effect that was also observed in
spring-block simulations~\cite{scheibert:2010,maegawa:2010}. The local
kinetic friction coefficient as well as the transition parameter were
not measured in the experiments. Therefore, they were studied here
qualitatively (see first paragraph of the following section) and
eventually chosen arbitrarily. When the ratio of the local tangential
traction to the contact pressure exceeds the static friction threshold
($t_s/p_s>\mu_s$), slip occurs and propagates in one or both
directions along the frictional interface. The dynamics of the slip
fronts are determined by the parameters of the friction law
(Eq.~\ref{eq:fric_law}) as well as by the local stress state.

\section*{Results \& Discussion}

We have conducted several simulations (not all presented in this
Letter) and have observed different types of slip: crack-like (the
entire interface between the crack tips is slipping), pulse-like (the
slip region propagates along the interface within a narrow pulse) and
mixed modes when a crack converts to pulses and vice versa.  The
propagation speed of the slip tip $V_{\mbox{\tiny{tip}}}$ is related
to the local stress state and seems not to depend on the type of slip.
By studying the influence of the friction law parameters, we have
observed that for an increasing (decreasing) difference between the
static and the kinetic friction coefficients $\Delta \mu = \mu_s -
\mu_k$, the slip type tends to be crack-like (pulse-like).  A higher
transition parameter $\alpha$ causes slower slip propagation
especially during slip initiation and slip arrest.

\begin{figure}
\begin{center}
 \includegraphics[width=0.7\textwidth]{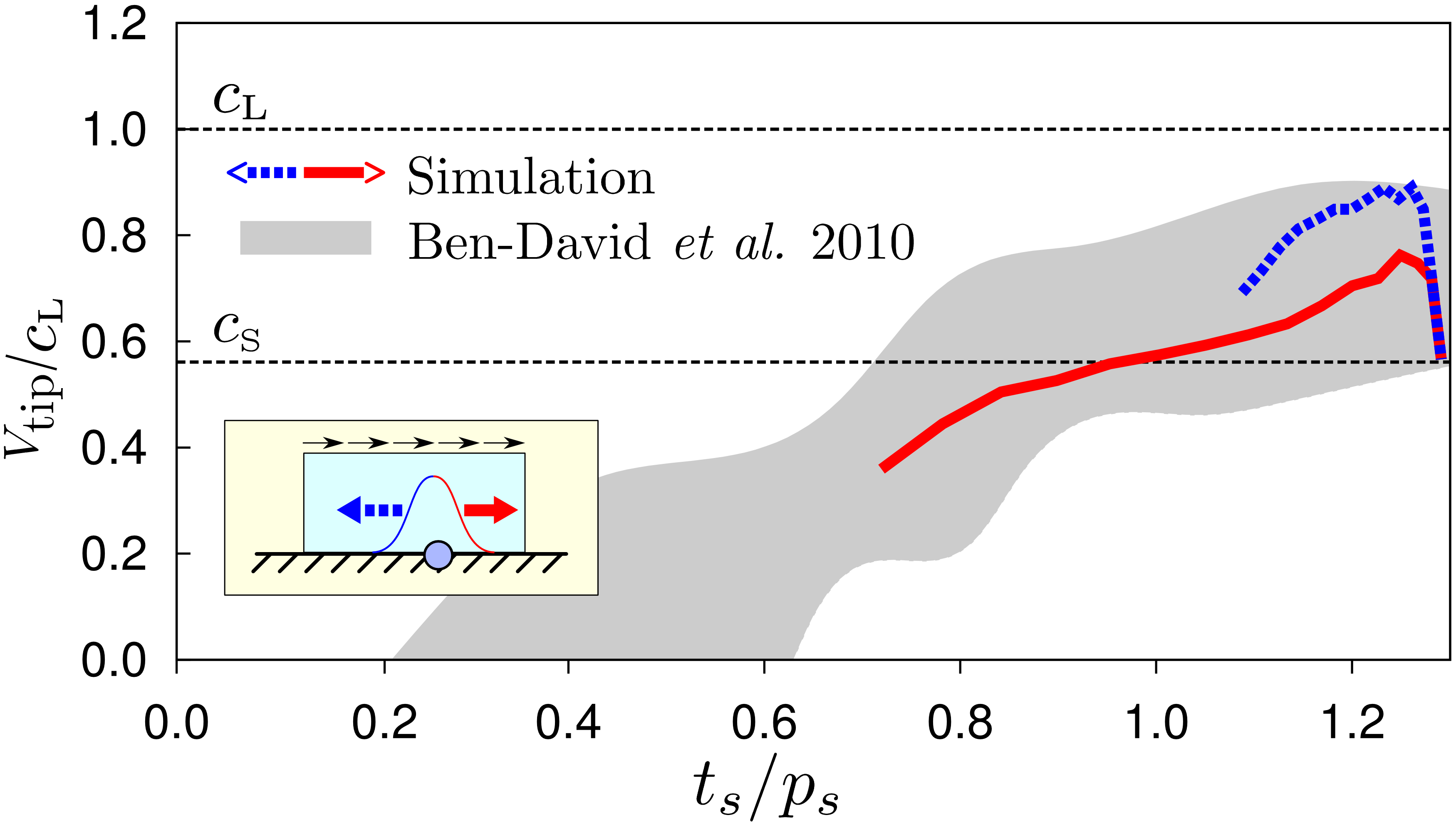}
 \caption{Comparison of numerical results with experimental observations
   by Ben-David \etal~\cite{ben-david:2010a}. The normalized rupture
   velocity is reported with respect to the static ratio of local
   tangential surface traction $t_s$ to contact pressure
   $p_s$. Friction parameters are $\mu_s = 1.3$, $\mu_k = 1.0$ and $\alpha =
0.1$\;m$^2$/s$^2$}
 \label{fig:fineberg}
\end{center}
\end{figure}

In order to compare our numerical results with the experimental
observations of Ben-David \etal~\cite{ben-david:2010a}, we present
$V_{\mbox{\tiny{tip}}}$ as a function of the ratio of shear to normal
stress measured before slip initiation. Here, the slip tip speed
$V_{\mbox{\tiny{tip}}}$ is normalized to the longitudinal wave speed
$c_{\mbox{\tiny{L}}}$ and the local stress ratio is replaced by the
local static ratio of tangential surface traction $t_s$ to contact
pressure $p_s$ [see Fig.~\ref{fig:fineberg}]. Our results confirm the
experimentally~\cite{ben-david:2010a} and
numerically~\cite{tromborg:2011} observed general trend that the
rupture propagation is faster for higher $t_s/p_s$ ratios.  For
friction parameters $\mu_s = 1.3$, $\mu_k = 1.0$ and $\alpha =
0.1$\;m$^2$/s$^2$ the slip front velocities are in good quantitative
agreement with the experimental results. Consistently with
experiments~\cite{ben-david:2010a}, for the given type of loading
(only at the top face), we do not observe slow fronts. Interestingly,
we note that the rupture propagates considerably slower in the
direction of the imposed shear load than in the opposite direction
[compare solid with dashed line in Fig.~\ref{fig:fineberg}]. These
differences have not been reported in the experiments.

\begin{figure} 
\begin{center}
 \includegraphics[width=0.7\textwidth]{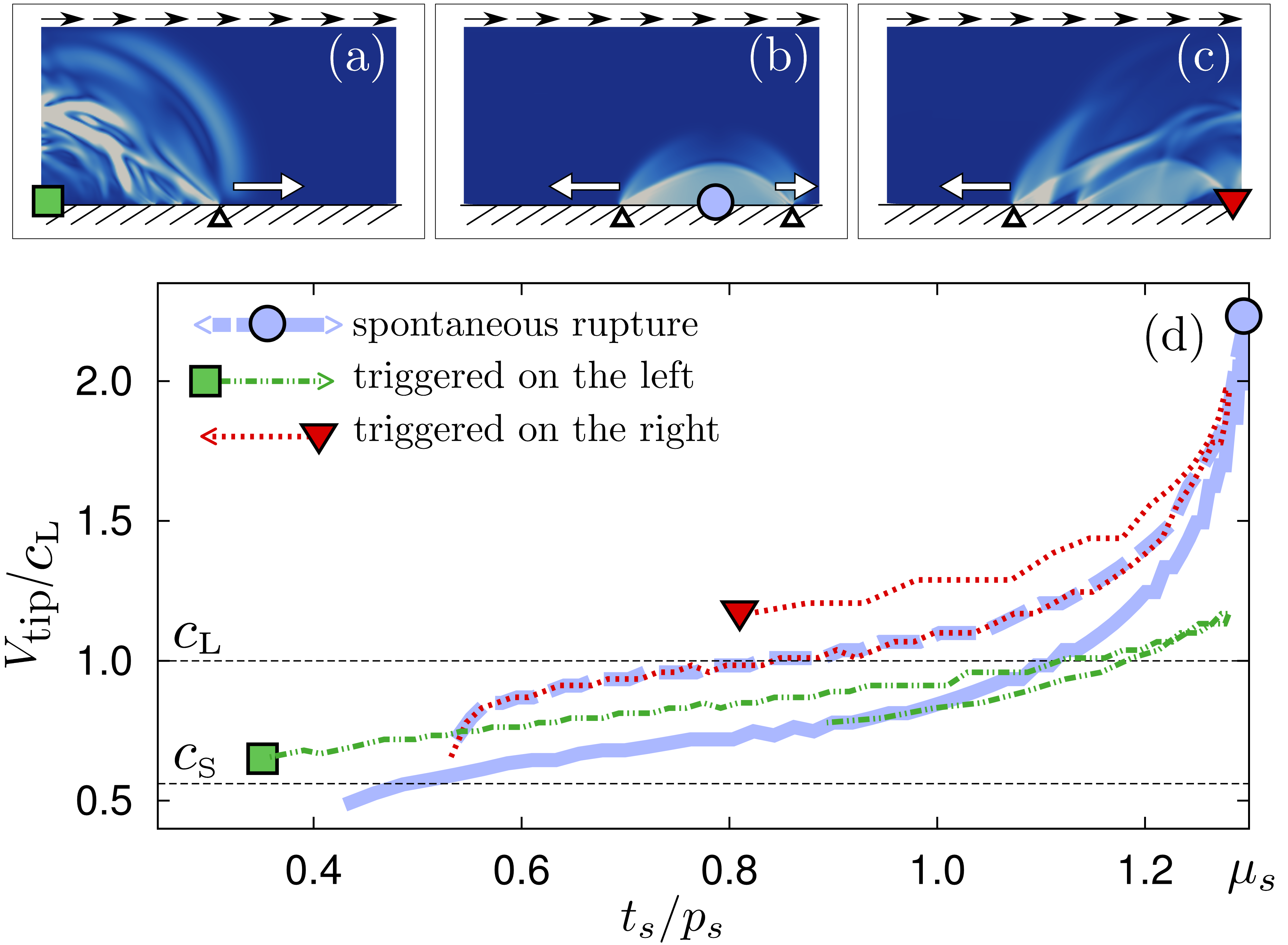}
 \caption{Three different slip events are presented for the same
   initial stress state (before triggering or spontaneous
   initiation). Instantaneous material velocity is shown for the slip
   event (a) triggered at the left edge, (b) spontaneously initiated
   far from the edges and (c) triggered at the right edge. Colors from
   blue to white denote material velocities ranging from $0$m/s to
   $2$m/s, respectively. The starting point of each event is marked
   with a square, a circle and a triangle, respectively. Small white
   triangles show the location of the tip of the slip front.  Black
   arrows indicate the direction of the imposed global shear load,
   whereas white arrows show the direction of the rupture propagation.
   (d) The normalized rupture velocity for all three cases is depicted
   with respect to the local static ratio of tangential traction $t_s$
   to contact pressure $p_s$ (data close to the triggering zone are
   not shown). Friction parameters are $\mu_s = 1.3$, $\mu_k = 0.6$
   and $\alpha = 0.1$\;m$^2$/s$^2$}
 \label{fig:old_criterion}
\end{center}
\end{figure}

\begin{figure} 
\begin{center}
 \includegraphics[width=0.7\textwidth]{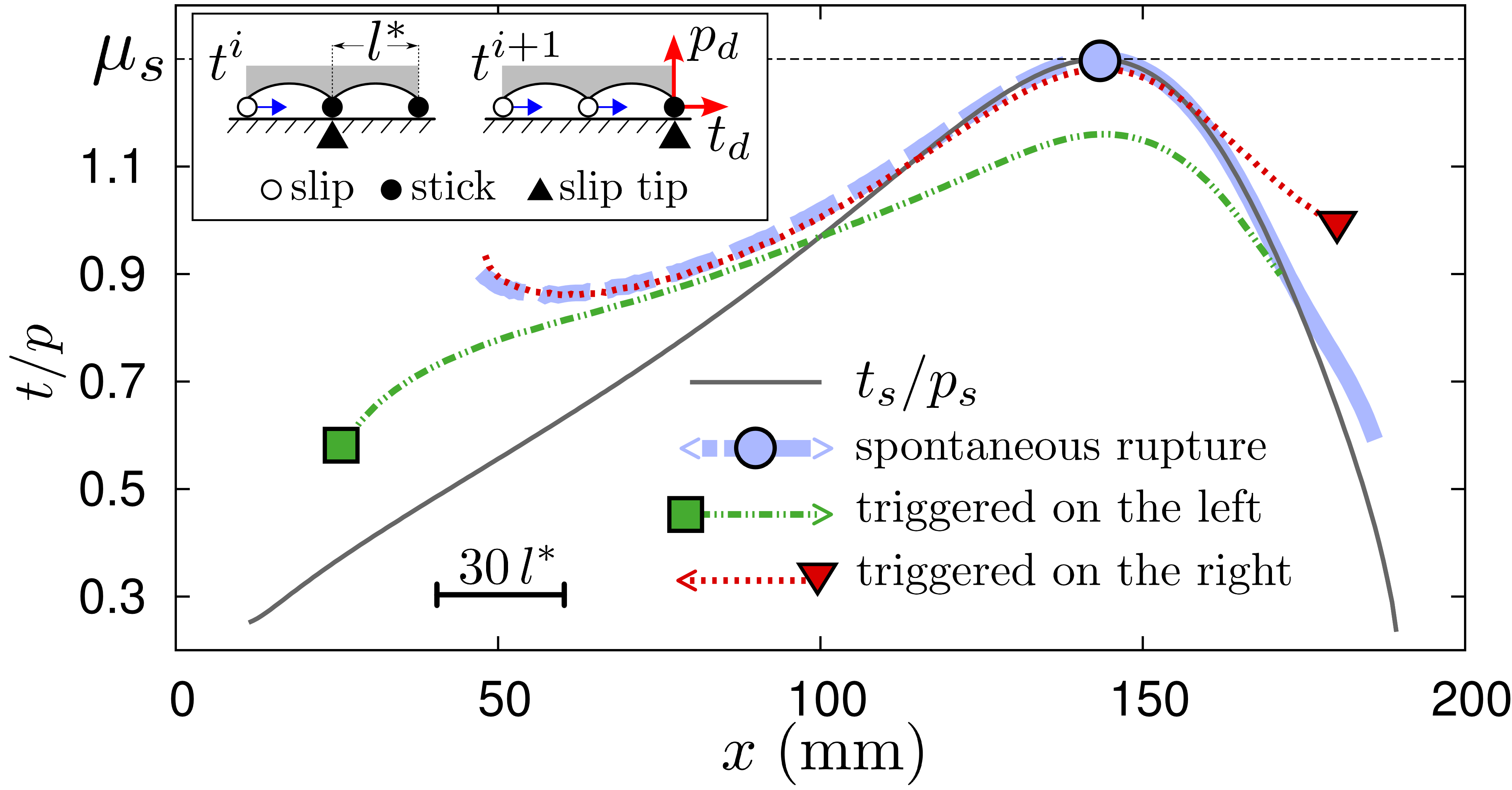}
 \caption{For each point along the interface $x$ the dynamic stress
   ratio $t_d/p_d$ is plotted at the moment when the slip front
   arrives at this location. Note that contrary to the reported static
   stress ratio $t_s/p_s$, this is not an instantaneous picture but
   an assembly of results over the entire time of propagation. Data
   close to the triggering zones are omitted. (Inset) The dynamic
   values $t_d$ and $p_d$ are measured at the sticking node in front
   of the slipping region at the moment $t^{i+1}$ when the previous
   node starts to slip}
 \label{fig:dynamic_t_p}
\end{center}
\end{figure}

To enable the separation of effects due to slip directionality and any
other sources that might cause a non-unique relation between the
$t_s/p_s$ ratio and the rupture propagation speed we consider two
additional simulations [Fig.~\ref{fig:old_criterion}], where slip
events are triggered at the edges. In order to increase the
propagation distance [in comparison to Fig.~\ref{fig:fineberg}, where
the rupture propagating in the opposite direction of the imposed shear
load arrests not far from the initiation zone] the kinetic coefficient
of friction is reduced resulting in the following set of friction
parameters: $\mu_s = 1.3$, $\mu_k = 0.6$ and $\alpha =
0.1$\;m$^2$/s$^2$.  In all three cases the loading history of the body
is identical up to the moment the tangential surface traction reaches
the friction threshold, \ie, the initial stress state is the same for
all simulations [see solid line in Fig.~\ref{fig:dynamic_t_p}].  The
slip propagation is then triggered by manually increasing the local
tangential surface traction within small nucleation zones at the edges
[Fig.~\ref{fig:old_criterion}(a),(c)]. Otherwise if the global shear
load is slightly increased, rupture nucleates spontaneously far from
the edges as before [Fig.~\ref{fig:old_criterion}(b)]. In case of
spontaneous initiation [Fig.~\ref{fig:old_criterion}(d), solid line]
the rupture propagates fast toward the edges and its velocity
decreases along the path with a decreasing ratio $t_s/p_s$. Note that
under some conditions we observe supersonic slip fronts, which were
not observed in~\cite{ben-david:2010a}. However, our results are
consistent with rupture in bi-material interfaces where the stiffer
material limits the propagation speed as observed experimentally and
numerically by Coker \etal~\cite{coker:2003}. For the two
edge-triggered ruptures the slip propagates relatively slowly in the
first phase, accelerates, reaches a maximum value (for maximal ratio
$t_s/p_s$) and decelerates afterwards [see
Fig.~\ref{fig:old_criterion}(d), dashed and dashed-dotted lines].
Although the triggered ruptures are unidirectional, there is no unique
slip tip speed associated with a given $t_s/p_s$ value.  The maximal
rupture velocity of the left-triggered slip does not exceed $60\%$ of
the maximal speed for the other two cases.

As seen most clearly in Fig.~\ref{fig:old_criterion}(a) the slip front
(marked by a small white triangle) propagating at super-shear velocity
follows the longitudinal wave (the circular white zone furthest from
the nucleation zone), which modifies the local stress state at the
interface. Therefore, looking at the dynamic ratio $t_d/p_d$ measured
in front of the slip tip, instead of examining the static ratio
$t_s/p_s$, would allow to account for the dynamic nature of the slip
propagation.

Here, the location of the slip tip is determined to coincide with the
position of the sticking node in front of the slipping nodes [see
inset in Fig.~\ref{fig:dynamic_t_p}].  According to this definition,
the position of the rupture tip changes abruptly when the front
advances. However, its velocity is computed in a continuous way as
$V_{\mbox{\tiny tip}} = l^*/\Delta t$, where $l^*$ is a characteristic
distance (here $l^* = 0.67\,$mm) and $\Delta t$ is the time interval
that the rupture needs to advance this distance.

In the context of discrete contact, we propose to analyze an
instantaneous dynamic stress state ($t_d$ and $p_d$) at the slip tip
right after it jumps to a new position [see
inset in Fig.~\ref{fig:dynamic_t_p}]. The dynamic ratio $t_d/p_d$
differs significantly from the static one [compared in
Fig.~\ref{fig:dynamic_t_p}], being changed by the longitudinal wave
often preceding the slip front. It is worth noting that the value of
the dynamic ratio is far from the critical value $\mu_s$ for a large
part of the propagation path, which implies the need for a strong
change of the stress state at the rupture tip within a short time.

\begin{figure} 
\begin{center}
 \includegraphics[width=0.7\textwidth]{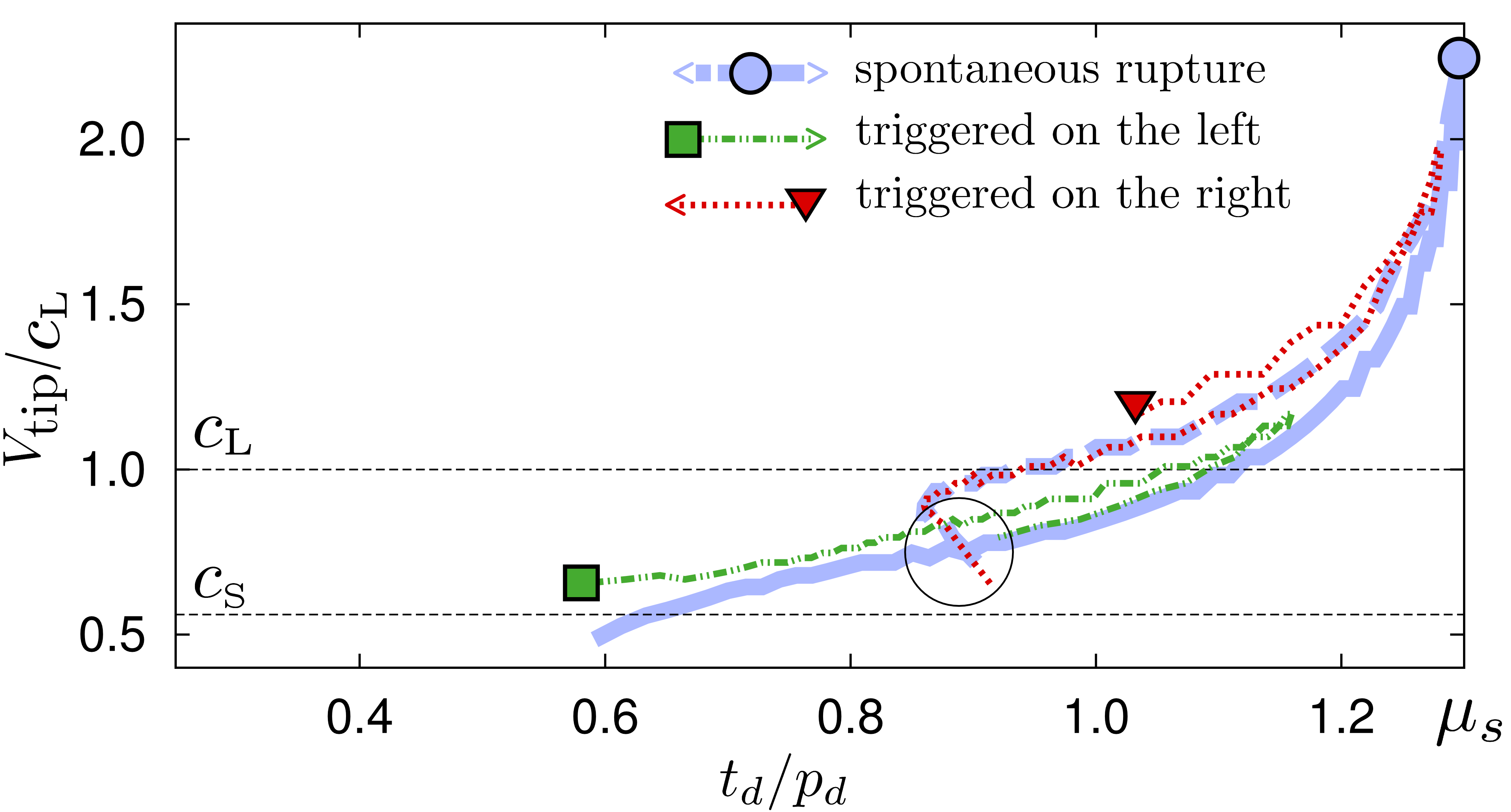}
 \caption{The normalized rupture velocity is plotted with respect to
   the dynamic traction ratio $t_d/p_d$ for all three slip
   events. Data close to the triggering zones are omitted}
 \label{fig:dynamic_criterion}
\end{center}
\end{figure}

The relation between the velocity of the slip front and the dynamic
ratio $t_d/p_d$ is depicted in
Fig.~\ref{fig:dynamic_criterion}. Compared to
Fig.~\ref{fig:old_criterion}, the rupture triggered on the left is in
better agreement with the other two (faster) slip
fronts. Particularly, the slopes are more consistent for all curves
and the range of velocities is smaller for a given ratio $t_d/p_d$.
Again it is confirmed that the character of the slip propagation is
directionality dependent. For a given ratio $t_d/p_d$, the slip fronts
propagating in the direction opposite to the sliding are faster than
the oncoming fronts [in Fig.~\ref{fig:dynamic_criterion}, \eg, compare
the dashed with the dashed-dotted curves]. Nonetheless, the difference
between the curves cannot be only attributed to the directionality [in
Fig.~\ref{fig:dynamic_criterion}, note the two branches of the dashed
and dashed-dotted curves]. The accelerating slip fronts show a faster
rupture velocity than the decelerating ones for the same given ration
$t_p/p_d$. Further, the general trend of faster rupture for higher
$t/p$ is lost [enclosed by the large circle in
Fig.~\ref{fig:dynamic_criterion}]; at a certain moment, the rupture
speed starts to decrease rapidly with increasing $t_d/p_d$ along the
propagation path.  We observe this phenomenon only for slip fronts
advancing against the sliding direction.  Regardless of the simplicity
of the static criterion $t_s/p_s$ and the consistency of the dynamic
criterion $t_d/p_d$, a stress ratio does not seem able to provide a
fully reliable estimation of the velocity of the slip propagation.

\begin{figure}
\begin{center}
 \includegraphics[width=0.7\textwidth]{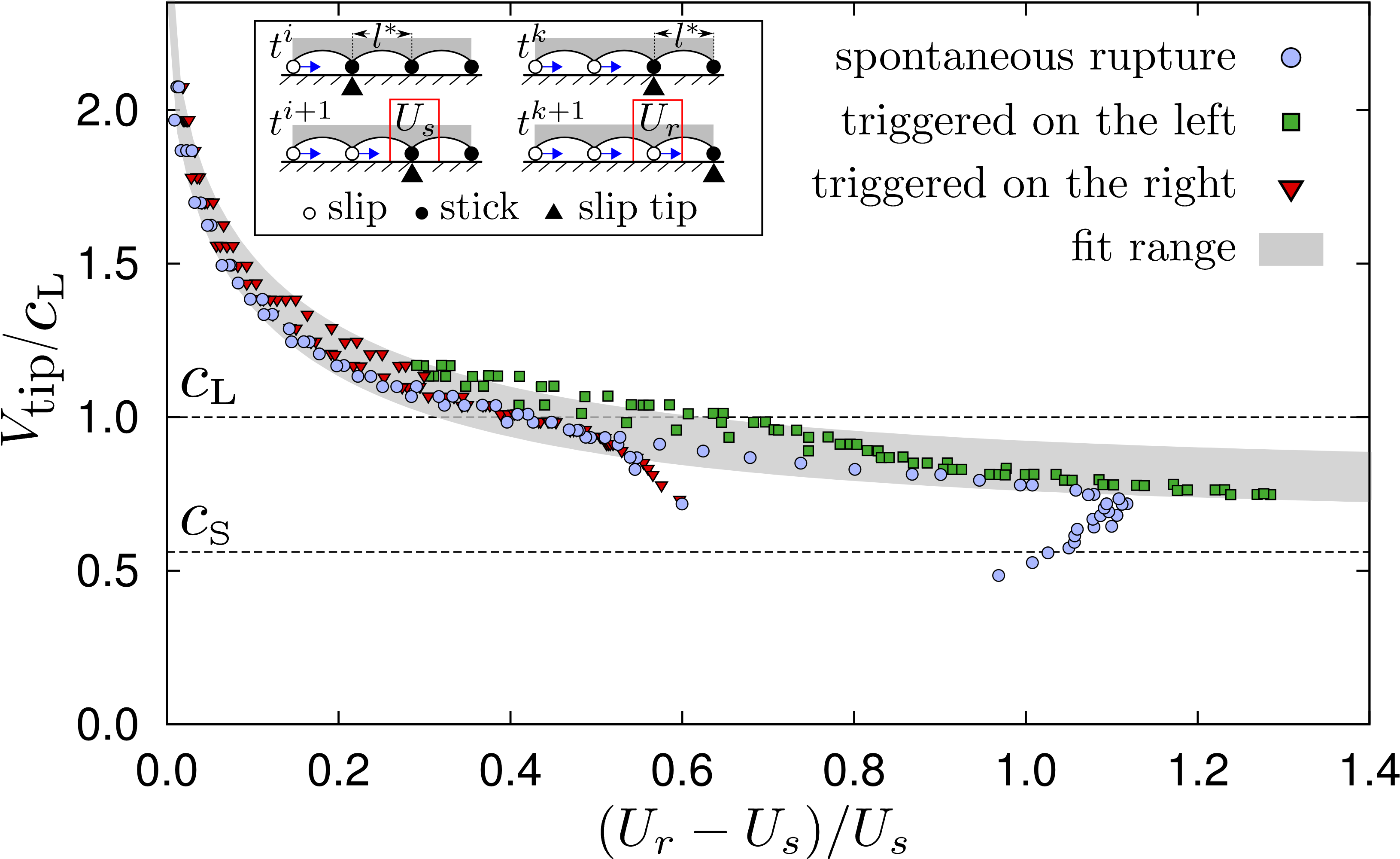}
 \caption{The normalized rupture velocity is depicted with respect to
   a dynamic criterion based on a heuristic surface energy density
   $U$. The ratio $\left(U_r - U_s\right)/U_s$ represents the
   proportion of the energy change $U_r - U_s$ at the slip tip needed
   to advance the rupture front with respect to the locally stored
   energy $U_s$. The gray area is a data fit based on
   Eq.~\ref{eq:fit_function} with $a=0.76\pm0.07$, $b=1.80$ and
   $c=3.05$. (Inset) $U_s$ and $U_r$ are measured when the slip tip,
   respectively, reaches the observation point and overpasses it}
 \label{fig:energy}
\end{center}
\end{figure}

The lack of generality of the velocity criteria based on the ratio of
the tangential traction to the contact pressure $t/p$ suggests an
independent consideration of $t$ and $p$. 
It was proposed~\cite{ben-david:2010a} that the propagation of the
slip front is related to the energy densities $U_s$, stored at the
front tip, and $U_r$, needed to advance the slip front. We propose a
heuristic energy density at the contact interface as
\begin{equation}
  U\left(p,t\right) = \left(2(1+\nu) t^2 + p^2\right)/2E~.
  \label{eq:heuristic_energy}
\end{equation}
The density of stored energy $U_s = U\left(p_d,t_d\right)$ is measured
locally at the slip tip at the moment the front advances one length
parameter $l^*$, similarly to the dynamic ratio $t_d/p_d$. The density
of rupture energy $U_r = U\left(p_r,\mu_s p_r\right)$ is computed at
the same material point just before the front advances another $l^*$,
\ie, when the ratio of tangential traction to contact pressure reaches
the static coefficient of friction ($t_r/p_r=\mu_s$) [see inset in
Fig.~\ref{fig:energy}].

The normalized rupture velocity is depicted in Fig.~\ref{fig:energy}
as a function of the change of the energy density at the slip tip
$\Delta U = U_r - U_s$ normalized by the stored energy density
$U_s$. The data of all three cases collapse within a narrow region
properly described by
\begin{equation}
  V_{\mbox{\tiny tip}} / c_{\mbox{\tiny{L}}} = a + b \exp(-c \sqrt{\Delta U / U_s})~,
  \label{eq:fit_function}
\end{equation}
where $a$, $b$ and $c$ are fitting parameters [see
Fig.~\ref{fig:energy}]. No differences due to the directionality of
the slip propagation nor any other reason that caused branching for
the previously studied criteria are now present. This shows that the
energy density criterion is able to account for the dynamics of slip
events at bi-material interfaces. Note that tails of data points
falling outside of the fit range occur when the slip fronts start to
decelerate rapidly before arresting.

\section*{Conclusion}

In this Letter it is demonstrated that the static ratio of shear to
normal stress~\cite{ben-david:2010a,tromborg:2011} is not a sufficient
criterion for determining the speed of slip fronts. The use of the
dynamic ratio, measured in front of the slip tip, improves the
estimation of this speed. However, for our set-up we observed that,
given a stress ratio (static or dynamic), the front going in the
direction of the sliding is always slower than the front propagating
in the opposite direction. Moreover, the decelerating fronts are also
slower than the accelerating ones. The energetic criterion we propose
eliminates these effects and highlights the similarities between the
rupture of frictional interfaces~\cite{coker:2005} and crack
propagation~\cite{coker:2003}. It is hoped that these findings
motivate experimental work to access dynamic stress field measurements
as well as theoretical studies to extend the principles of fracture
mechanics to problems of frictional sliding.

\section*{Acknowledgments}

We are grateful to D. Coker for fruitful discussions and N. Richart
for helpful advices on the simulations. The research described in this
Letter is supported by the European Research Council (ERCstg
UFO-240332).

\bibliographystyle{tribolett}

\newpage

\begin{center}
\begin{table}
  \caption{Friction and material parameters corresponding to polymethylmethacrylate glass, PMMA}
  \label{tab:material}
  \begin{center}
  \begin{tabular}{  l  r l }
    \hline
    Parameter &  & \\
    \hline
    \textit{Material:} & & \\
    \hspace{0.1cm} Young's modulus $E$ & 2.6 & GPa \\
    \hspace{0.1cm} Poisson's ratio $\nu$ & 0.37 & \\
    \hspace{0.1cm} Density $\rho$ & 1200 & kg/m$^3$ \\
    \hspace{0.1cm} Longitudinal wave speed $c_{\mbox{\tiny{L}}}$ & 1584 & m/s \\
    \hspace{0.1cm} Transverse wave speed $c_{\mbox{\tiny{S}}}$ & 890 & m/s \\
    \textit{Friction:} & & \\
    \hspace{0.1cm} Static friction coefficient $\mu_s$ & 1.3 & \\
    \hspace{0.1cm} Kinetic friction coefficient $\mu_k$ & \hspace{1cm}0.6;\;1.0 & \\
    \hspace{0.1cm} Transition parameter  $\alpha$ & 0.1 & m$^2$/s$^2$ \\
    \hline  
  \end{tabular}
  \end{center}
\end{table}
\end{center}

\end{document}